\documentclass[aps,prl,twocolumn,showpacs,groupedaddress]{revtex4-1}

\usepackage{graphicx}
\usepackage{dcolumn}
\usepackage{bm}
\usepackage{amsmath}
 \usepackage{array}

\begin{document}

\title{Resonant magneto-acoustic  switching:
influence of Rayleigh wave frequency and wavevector}

\author{P. Kuszewski$^{1}$, I. S. Camara$^{1}$,  N. Biarrotte$^{1}$, L. Becerra$^{1}$, J. von Bardeleben$^{1}$, A. Lema\^itre$^{2}$, C. Gourdon$^{1}$,   J.-Y. Duquesne$^{1}$, and L. Thevenard$^{1}$\email[e-mail: ]{thevenard@insp.jussieu.fr}}

 
\affiliation{
$^1$ Sorbonne Universit\'e, CNRS,  Institut des Nanosciences de Paris, 4 place Jussieu,75252 Paris France\\
$^2$  Centre de Nanosciences et de Nanotechnologies, CNRS, Univ. Paris-Sud, Universit\'e Paris-Saclay, 91460 Marcoussis, France\\
}

\date{\today}

\label{sec:Abstract}

\begin{abstract}


 We show on in-plane magnetized   thin films that  magnetization can be switched efficiently by 180 degrees using large amplitude Rayleigh waves travelling along the hard or easy magnetic axis.  Large characteristic filament-like domains are formed in the latter case. Micromagnetic simulations clearly confirm that  this   multi-domain configuration  is compatible with a resonant precessional mechanism.    The reversed domains are   in both geometries  several hundreds of $\mu m^{2}$,  
 much larger than hasàç been shown using spin transfer torque- or field-driven precessional switching. We show that surface acoustic waves can travel at least 1mm before addressing a given area, and can interfere to create magnetic stripes that can be positionned with a sub-micronic precision.

\end{abstract}



\maketitle


\section{Introduction}

    Resonant magnetization switching relies on triggering large angle magnetization precession by high frequency stimuli.  Sub nanosecond  deterministic switching was thus demonstrated using short magnetic field pulses\cite{Back1999,Choi2001,Gerrits2002,Hiebert2002,Schumacher03,Devolder2008,Maunoury2005},   spin transfer torque  (STT) \cite{Krivorotov2005,Papusoi2009,Devolder2005,Vaysset2011,Zhang2012,Bedau2010}, electric fields or ultra-short light pulses\cite{Shiota2011,Ghidini2013a,Stupakiewicz2017} to induce an efficient torque on the magnetization.  An alternative  route  relies on the inverse-magnetostriction from  acoustic waves\cite{Thevenard14,Gowtham2016,Dreher12}. In the right conditions\cite{Kovalenko2013,Thevenard2013,Thevenard2016}, they could drive  resonant precessional reversal, with the noteworthy benefit  that their low attenuation should enable an action  remote from their generation, with wave-front shaping or stationary wave excitation to  adress selectivally a given area\cite{Li2014}. Voltage-driven "straintronics" also hold the promise of a lower power consumption\cite{Balestriere2010,Shiota2011,Ghidini2013a}.




   

In this framework, we report on   the optimum conditions for  acoustic magnetization reversal, and also hint to its limits. For this study, we chose to work on a dilute magnetic semiconductor, (Ga,Mn)As, in which the carrier-mediated ferromagnetic phase exhibits magnetoelasticity. Although its Curie temperature remains well below room-temperature, the tunability of its magnetic properties has proven instrumental to  demonstrate key concepts in spintronics\cite{Elsen2006a,Bihler2008,Cormier2014,Thevenard2016}. 

Here we present a   study of the influence of the  travelling/stationary SAW direction with respect to the magnetic easy axis and of the SAW frequency up to about 1GHz. We show how these govern the efficiency of the switching, but also the size and shape of the domains.  In particular,   homogeneous domains several hundreds of microns wide are obtained on in-plane magnetized (Ga,Mn)As, a substantial improvement over out-of-plane magnetized (Ga,Mn)(As,P) samples\cite{Thevenard2016} in which only sub-micron domains were obtained. Finally, micromagnetic simulations validate the  resonant switching mechanism at work, and the shape of the domains.

	\begin{figure}
	\centering
	\includegraphics[width=0.49\textwidth]{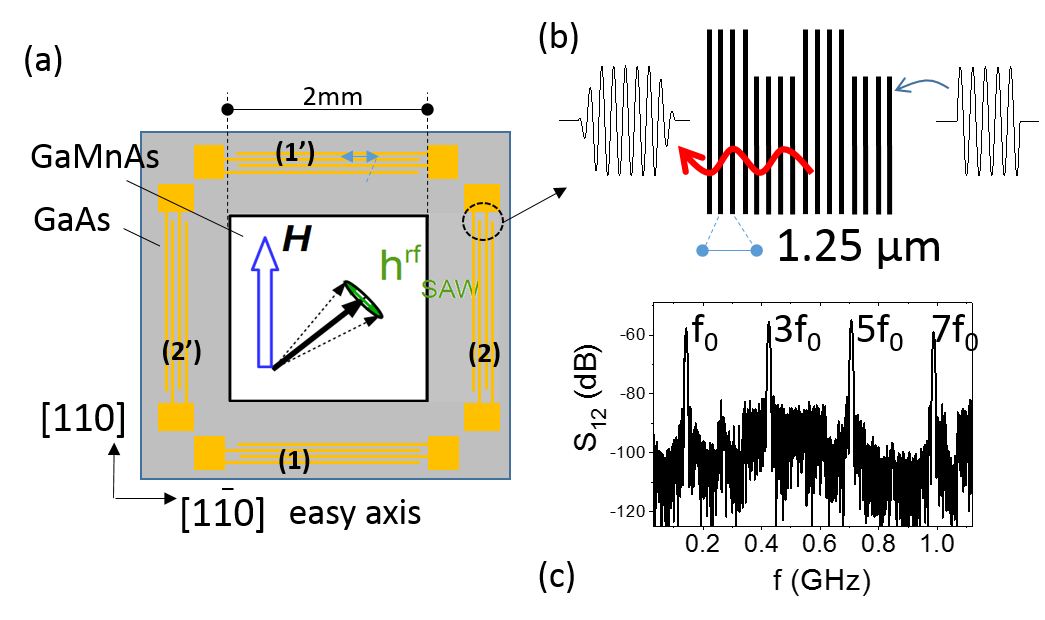}
	\caption{(a) Schematics [not to scale]: exciting  IDT \textbf{2} (resp. \textbf{1}) generates a SAW travelling along  [-110] (resp. [110]). In the SAW-FMR regime, a small angle precession of the magnetization (black arrow) ensues, provided a hard-axis  field is applied (blue arrow). (b) Schematics of the "split 44" design of the IDT. An rf pulse excites the transducer which generates a strain pulse. (c)  Gated transmission $S_{12}$ measured using a Vector Network Analyzer ($T$=20K, $\vec{k}_{saw}$//[1-10]).}
	\label{fig:SAWFMR}
\end{figure}


\section{Experimental methods}

The sample is a $h$=45-nm-thick layer of (Ga,Mn)As grown by molecular beam epitaxy on non intentionnally doped (001) GaAs. 
A  16h/200$^{\circ}$C anneal results in a  Curie temperature of T$_{C}$=120~K and a magnetically active Mn concentration of  $x$=5$\%$. The layer is magnetized in-plane   with a   uniaxial anisotropy along [1-10]. The  cubic components of the anisotropy are   much weaker than the uniaxial terms \cite{Linnik2011,Thevenard14}.

	\begin{figure}
	\centering
	\includegraphics[width=0.49\textwidth]{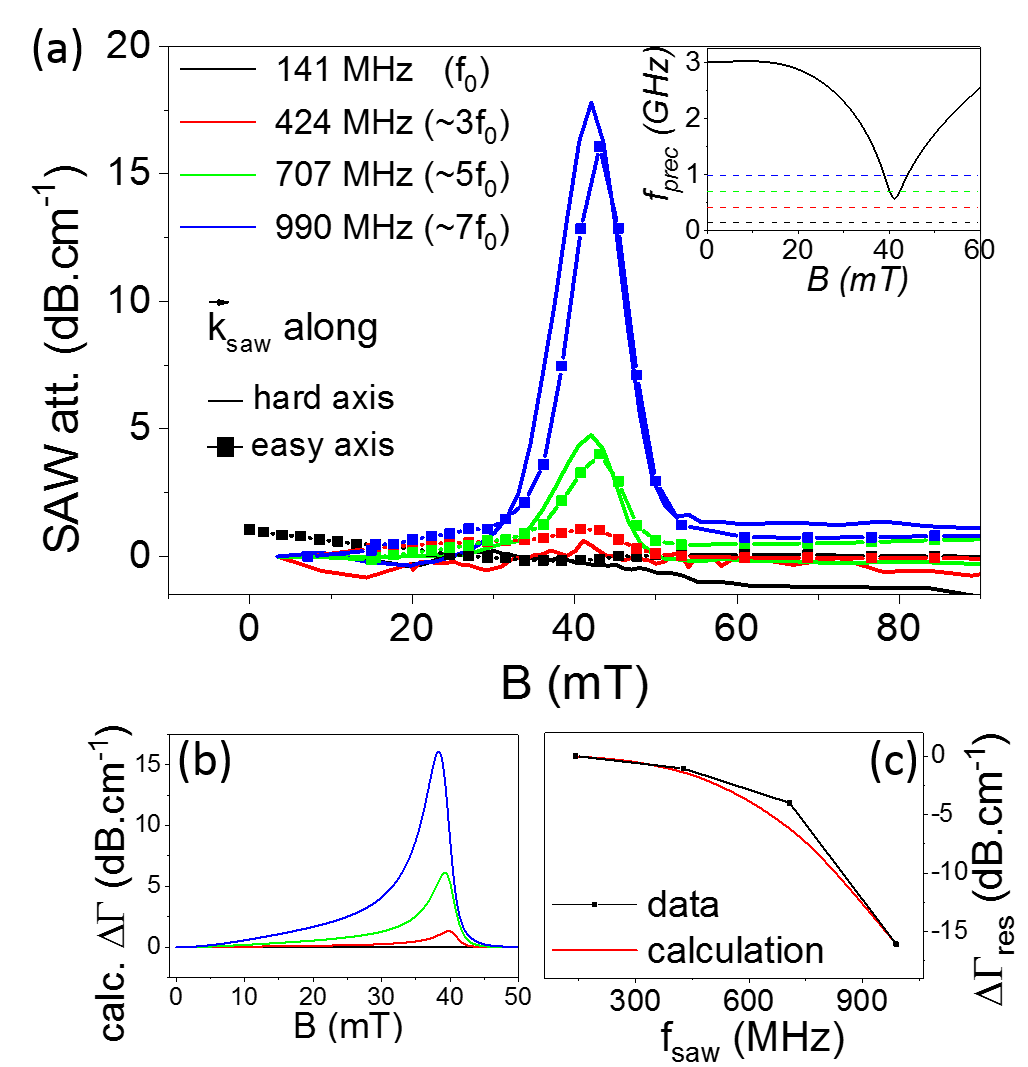}
	\caption{SAW-FMR regime: (a) Acoustic attenuation variations for a  SAW propagating along the easy ([1-10]) or  hard ([110]) axis, for 4 different frequencies ($T$=20K). The acoustic power was 16mW for $5f_0$ and $7f_0$, 50mW for $3f_0$ and 85mW for $f_0$ (the different powers reflect  the frequency dependency of the excitation electronic circuit).  Up to these  powers, corresponding to a surface strain of $\varepsilon_{xx}\approx  10^{-5}$, we observe that the attenuation and velocity variations are power independent (linear regime).  The inset is a schematics of the field-dependence of the precession frequency. The resonance field lies where  f$_{prec}(B)$ is closest to the the first intersection with the SAW frequency (dotted lines). At the second crossing,   the effective rf field generated by the SAW is almost zero \cite{Thevenard14}. (b) Acoustic attenuation variations calculated for a  SAW propagating along  [1-10] at  $f_0$, $3f_0$, $5f_0$ and  $7f_0$ -  same color coding as (a).  (c) SAW attenuation variations at resonance, extracted from the data of (a) and the calculation of (b).}
	\label{fig:SAWFMReffetf}
\end{figure}

The SAWs were excited electrically, using interdigitated transducers (IDTs). A 2$\times$2mm$^2$ (Ga,Mn)As mesa was first defined by wet etching (figure \ref{fig:SAWFMR}a). Four IDTs of 1mm aperture were then evaporated by a   60nm-thick Al lift-off, using the "split-44" transducer design \cite{Schulein2015} with 15 pairs of 4 equi-potential digits  and a nominal base periodicity of $\lambda$=20$\mu m$  (figure \ref{fig:SAWFMR}b). Relying on the natural piezoelectricity of GaAs, surface acoustic waves travelling along [110] (resp.[1-10]) were  excited using IDTs \textbf{1} (resp. \textbf{2}), inducing 2 longitudinal strain components\footnote{There is also a shear  $\varepsilon_{xz}(t)$ component. But it is close to zero  at the surface.  Within the layer, it is negligeable since the layer is located just below the surface and is very thin with respect to the smallest penetration depth ($\lambda_{7f_0}$=2.9$\mu m$.}: $\varepsilon_{zz}(t)$ and $\varepsilon_{xx}(t)$, with $x$//$\vec{k}_{saw}$ and $z$  normal to the sample plane. After propagating across the (Ga,Mn)As mesa, they were detected by IDTs  \textbf{1'} (resp. \textbf{2'}). The frequency of the SAW was tuned to $nf_0$  with $n$=1,3,5,7 and $f_0$=$\frac{V_r}{\lambda}\approx$141 MHz,  where $V_r$ is the temperature-dependent Rayleigh velocity  (figure  \ref{fig:SAWFMR}c).   This design enabled us to study the influence of both the direction and the frequency of the SAW on the magnetization reversal.

Since the zero-field precession frequency of the layer - a few GHz as estimated from the anisotropy constants measured by cavity ferromagnetic resonance (FMR) - was systematically larger than the  SAW frequencies,  a field was  applied along the hard axis [110]  using an air-cooled CAYLAR dipole to bring the magnetization precession frequency down into resonance with the SAW (inset of figure \ref{fig:SAWFMReffetf}a). 

\section{SAW-FMR regime}

Before using high amplitude SAWs to switch the  magnetization in (Ga,Mn)As, we show that  acoustic waves can  excite FMR equally  whether  propagating along the hard or the easy magnetic axes.  SAW-FMR shows up as a resonant attenuation of the wave amplitude or a resonant velocity change\cite{Thevenard14,Gowtham2016,Dreher12,Labanowski2016}. To measure them, the layer is first magnetized uniformly along the easy [1-10] axis and 400ns-long rf bursts are sent on the exciting IDT (\textbf{1} or \textbf{2})  at 1 kHz repetition frequency. The amplitude/phase of the transmitted pulse are then monitored with respect to the hard-axis field along [110] (see references \onlinecite{Thevenard14,Thevenard2016} for a more detailed description of this time-domain SAW-FMR measurement  procedure). Amplitude and phase variations were measured versus field at the  four  SAW frequencies for the two orthogonal travelling directions, and  a large range of powers.  For the sake of conciseness we will focus here on the low power 20K attenuation data (figure \ref{fig:SAWFMReffetf}a),  sufficient to convey our message. The amplitude and velocity variations  measured at other temperatures and powers were quite similar qualitatively to the data shown in Refs. \onlinecite{Thevenard14,Thevenard2016}.

 A  pronounced resonance is evident in figure \ref{fig:SAWFMReffetf}a, as had already been seen in out-of-plane magnetized (Ga,Mn)(As,P)\cite{Thevenard14} or in-plane magnetized Nickel\cite{Dreher12,Gowtham2015,Labanowski2016}: the SAW triggers   small oscillations of the magnetization  around its equilibrium position (schematics in figure \ref{fig:SAWFMR}a).   Using the cavity-FMR determined anisotropy constants, we calculate  f$_{prec}(B)$ (inset of figure \ref{fig:SAWFMReffetf}a) and see that the experimental resonance field  is  around 40mT,  for which the precession frequency stands closest to the SAW frequencies. The resonance is close to the   saturation, where the f$_{prec}(B)$ curve has a steep tangent. This explains the weak variation of the experimental resonance position with f$_{saw}$. Note that a 0.2$^\circ$ misalignement of the field with respect to the [110] axis was taken in the calculation, which will be justified in the following.

The amplitude and position of the acoustic resonance are very similar for SAWs propagating along  the hard  or easy  axis. This can be explained by computing  the effective field generated by the SAW. In the reference frame of the equilibrium magnetization\cite{Dreher12,Thevenard14}  $\mu_{0}\vec{h}_{rf}(t)$=(0,$\mp\frac{1}{2}A_{2xy}\varepsilon_{xx}(t)sin2\varphi_{0}+A_{4\varepsilon}[\varepsilon_{zz}(t)-\frac{\varepsilon_{xx}(t)}{2}]sin4\varphi_{0}$), where  $x$// $\vec{k}_{saw}//[1\pm10]$. The position of the static  magnetization, $\varphi_{0}$, is defined with respect to [1-10]. Here we have introduced the uniaxial and biaxial magnetostrictive coefficients  $A_{2\varepsilon},A_{4\varepsilon}$ and $A_{2xy}$ \cite{Linnik2011}:  $K_{2||}$=$\varepsilon_{XY,0}A_{2xy}M_S$ and  $K_{2\bot}$=$ -M_S(A_{2\varepsilon}-2A_{4\varepsilon})(\varepsilon_{ZZ,0}-\varepsilon_{XX,0})+\frac{\varepsilon_{XY,0}A_{2xy}M_S}{2}$. $K_{2||}$, $K_{2\bot}$ are the out-of-plane and in-plane uniaxial anisotropies, $M_S$ is the magnetization at saturation, and $\varepsilon_{XX,0}$, $\varepsilon_{ZZ,0}$ and $ \varepsilon_{XY,0}$  are the static biaxial and shear strains (in the GaAs $<$100$>$ axes frame). Because $A_{2xy}\gg A_{4\varepsilon}$ in this material\cite{Linnik2011,Thevenard2013}, the $A_{4\varepsilon}$ term in  $\mu_{0}\vec{h}_{rf}(t)$ is several thousand times smaller than the $A_{2xy}$ term, so the absolute value of the tickle field has a  very similar field  dependence (via $\varphi_{0}(B)$) for the two orthogonal SAW wave-vectors $\vec{k}_{saw}//[1\pm10]$.


 Finally, the attenuation variations at resonance increase with the SAW frequency, from barely detectable at $f_0$ to over 15 dB$.cm^{-1}$ at $7f_0$=990 MHz (figures  \ref{fig:SAWFMReffetf}a,c). This is a standard SAW FMR behavior since the absorbed acoustic power  varies as $P_{abs}$=-$\frac{\omega_{SAW}}{2}Im (\mathbf{h_{rf}^{*}}[\mathbf{\chi}]\mathbf{h_{rf}})$ where  $\chi$ is the magnetic susceptilibity tensor\cite{Dreher12,Gowtham2015}. Using the semi-infinite model developed for (Ga,Mn)(As,P) in Ref. \cite{Thevenard14}, we calculate the  field-dependence of the  SAW attenuation variations, assuming  a SAW frequency dependent filling factor $F(f_{saw})$=0.78$\frac{h}{\lambda_{saw}}$,  a static shear strain $\varepsilon_{XY,0}$=2.10$^{-4}$  and a 0.2$^\circ$ misalignement of the field with respect to the [110] axis (figure  \ref{fig:SAWFMReffetf}b). The shape and amplitude of the experimental SAW attenuation are well reproduced, as well as the frequency dependence of the attenuation at resonance (figure  \ref{fig:SAWFMReffetf}c). This strong dependence of the SAW attenuation with SAW frequency exists in both the linear and non-linear magnetization dynamics regimes, and  will be critical for the switching efficiency.

		 \begin{figure*}
	\centering
	\includegraphics[width=0.98\textwidth]{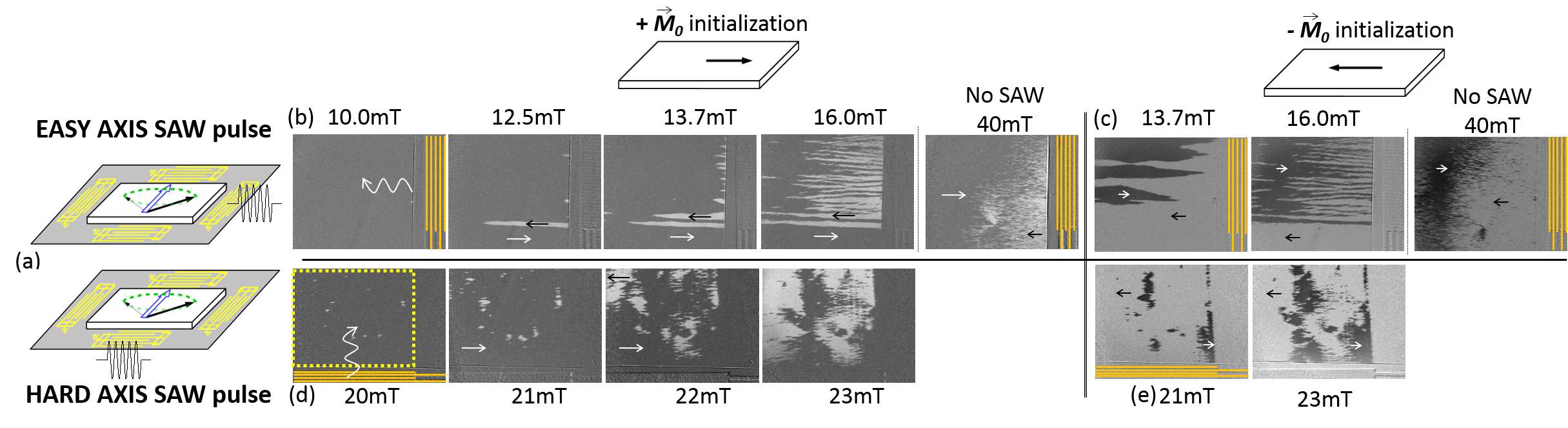}
	\caption{SAW switching regime. (a) Schematics [not to scale]. At large rf power, the SAW  triggers large angle precession of the magnetization (black arrow) around the applied field (blue arrow).  (b-e) Normalized Kerr images showing the result of a single $f_{saw}$=990 MHz 200ns-long rf burst on the uniformly magnetized layer (image 306$\times$410~$\mu m^2$). (b,c) $\vec{k}_{saw}$//[1-10]  (easy axis). Control images show what happens without applying any SAW pulse. (d,e) $\vec{k}_{saw}$//[110] (hard axis). (b,d) After  a \textbf{+$\vec{M}_0$} initialization, switched areas appear white. (c,e)  After a \textbf{-$\vec{M}_0$} initialization, switched areas appear black.}
	\label{fig:Switch2configs}
\end{figure*}

 \section{SAW-induced resonant switching}

 To observe the effect of a single SAW burst on the magnetization,  a longitudinal Kerr microscope, sensitive to the [1-10] component of the magnetization, was used. More details on this  set-up can be found in reference \onlinecite{Thevenard12}. Normalized Kerr images are taken as follows. The sample is first initialized magnetically $+\vec{\textbf{M}}_0$ or $-\vec{\textbf{M}}_0$ along the easy axis and a reference image  taken.  Applying a constant  magnetic field along [110], a \textit{single} SAW burst is then emitted  (figure  \ref{fig:Switch2configs}a), chosing its duration $\tau_{saw}$ to ensure  a  plateau with constant strain wave amplitude (the   rise/decay time of the acoustic burst is rather long - about 85ns - due to the transient passage of the wave through the excitation  transducer digits, see figure \ref{fig:SAWFMR}b).   After bringing the field   back to zero, a second  Kerr image is taken, and divided by the reference one. The sample was reinitialized before any new SAW pulse was applied. Particular care was given to align the  static bias field along the hard axis to impede  domain nucleation, which could compete with the precessional reversal (hysteresis cycles are very square, with a coercive field of 2 mT at 20 K). Finally, as previously discussed and quantified in references \cite{Thevenard2016,ThevenardNuc16} since no current is applied to the system, and thermoelasticity of GaAs is weak, no substantial temperature rise of the sample is expected from the passage of the SAW.


\subsection{Influence of the SAW direction on the shape of switched domains}

We first  show data for a SAW burst travelling along the easy [1-10] axis after a $+\vec{\textbf{M}}_0$ initialization.  Figure \ref{fig:Switch2configs}b shows normalized images taken close to the exciting IDT \textbf{2} after a single 200ns-long  burst ($7f_0$=990 MHz, $P_{exc}$=3.5W). This generated a  surface strain component  $ \varepsilon_{xx}$(z=0)$\approx$2$\times 10^{-4}$ (see Appendix A for how the SAW power and amplitude are estimated from $P_{exc}$ - in the following we will drop the mention "z=0").  The  reversal starts at low field, 12.5mT, and at the lower edge edge of the SAW wave-front where the  displacement is higher due to the finite aperture of the transducer\cite{royer2000elastic}. As the field is increased, the switching proceeds along filaments parallel to the SAW wave-vector. At 16mT and above, the magnetic pattern does not change anymore and is made of alternating $+\vec{\textbf{M}}_0$ and $-\vec{\textbf{M}}_0$  magnetized strips about 10-20$\mu m$ wide, and several hundreds of microns long. The same experiment was then done with the SAW burst propagating along the hard magnetic axis [110] this time, parallel to the static bias field. Figure \ref{fig:Switch2configs}d shows images for the same SAW frequency as above, taken close to the exciting  IDT \textbf{1}, after a single SAW burst ($P_{exc}$=1.5W). The  estimated  SAW amplitude is a little bit smaller than above, $ \varepsilon_{xx}\approx$7$\times 10^{-5}$, so that switching starts at a slightly higher field, around 20 mT. The reversed domains have a very different shape. They now form patches or  columns parallel to the wave-vector, with jagged edges.  At 23mT, domains several hundreds of $\mu m^{2}$ have reversed on the SAW path.

  In both  configurations, the switching pattern does not depend on the SAW pulse duration, as long as it is longer than the transient regime. For shorter pulses,   reversal persists in the (magnetically) weaker parts of the sample. It is observed at all temperatures up to $T_{C}$ at decreasing fields, as expected from the temperature dependence of the resonance \cite{Thevenard14}. No reversal was observed out of the SAW path, or when the IDT was excited out of its resonance.   Monitoring the SAW amplitude versus field in both configurations shows that it now resonates  lower than the 40 mT   observed in the linear regime (figure \ref{fig:SAWFMReffetf}a):  25 mT for $P_{exc}$=3.5W when  $\vec{k}_{saw}$//[1-10] and 30mT for  $P_{exc}$=1.5W  when $\vec{k}_{saw}$//[110]. This power-dependent downshift of the resonance  had already been observed in out-of-plane magnetized (Ga,Mn)(As,P)\cite{Thevenard2016} and is expected in the non-linear regime of magnetization precession, regardless of what drives it, be it an rf field\cite{Gnatzig1987} or spin transfer torque\cite{Chen2009b}.

SAW switching also works after a $-\vec{\textbf{M}}_0$ initialization (figures \ref{fig:Switch2configs}c,e). At low field, although the switched domains have the same overall shape, they are not located in the same region of the sample nor cover the same area as when starting from  $+\vec{\textbf{M}}_0$  (see the images at 13.7 mT for $\vec{k}_{saw}$//[1-10] and 21mT for $\vec{k}_{saw}$//[110] in figures \ref{fig:Switch2configs}b,d). Likewise,  only about half the observed zone has switched on average at  high field  for  $\vec{k}_{saw}$//[1$\pm$10]. The final switching pattern starting from +$\vec{\textbf{M}}_0$  is  complementary to the one starting from -$\vec{\textbf{M}}_0$:  only the areas unswitched starting from  +$\vec{\textbf{M}}_0$  have switched. The unfortunate consequence of this is that magnetization cannot be toggled from $\pm\vec{\textbf{M}}_0$ to $\mp\vec{\textbf{M}}_0$ by long SAW pulses. A possible explanation for this is a weak variation of the precise alignment of the field along [110] across the image.

\begin{figure}
	\centering
	\includegraphics[width=0.48\textwidth]{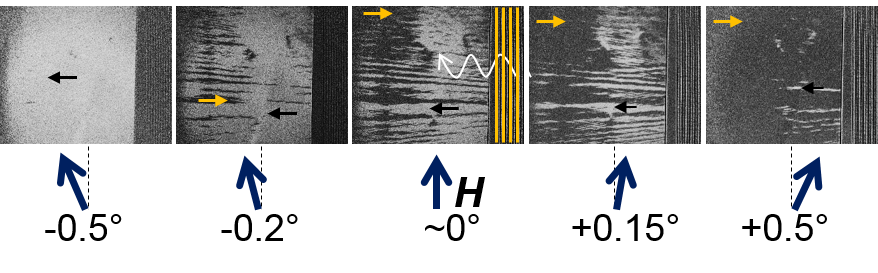}
	\caption{Final magnetization position after a 400-ns long pulse, depending on the misalignment of the magnetic coil   with the [110] direction, taken as the 0$^{\circ}$ position (31mT bias field, $T$=20K, $\vec{k}_{saw}$//[1-10],  images 306$\times$410~$\mu m^2$).}
	\label{fig:tilt}
\end{figure}

 To test the latter, we first do a control  experiment where the magnetization is initialized +$\vec{\textbf{M}}_0$ or -$\vec{\textbf{M}}_0$, the field ramped up to 40mT, and then brought back down to zero, \textit{without having applied any SAW pulse} (figures \ref{fig:Switch2configs}b,c, "NO SAW" images). If the magnetization is brought fully along [110] under this field, we expect a collection of equiprobable +$\vec{\textbf{M}}_0$ and -$\vec{\textbf{M}}_0$ domains at remanence.  The normalized images clearly show that this only occurs on one part of the image or another, depending on the initial magnetization, and that these regions correspond precisely to those switching under SAW. This could be well be explained by the existence of slightly curved field lines : if the sample is not exactly half-way between the coil poles, the field could be slightly offset to the right of [110] on the left part of the image, and  to the left of [110] on the right part.  Secondly,   the angle of the coil was varied in a controlled fashion. The resulting   percentage of averaged switched magnetization  goes from 50$\%$ to  100$\%$ with a disalignment of  a fraction of a degree   towards the $+\vec{\textbf{M}}_0$ or $-\vec{\textbf{M}}_0$ directions (figure \ref{fig:tilt}). Hence SAW-driven resonant switching is extremely sensitive to the precise alignment of the magnetic coil, as already shown experimentally and numerically on out-of-plane (Ga,Mn)(As,P)\cite{Thevenard2016}

\subsection{Influence of the SAW frequency on the switching efficiency}

		 \begin{figure}
	\centering
	\includegraphics[width=0.49\textwidth]{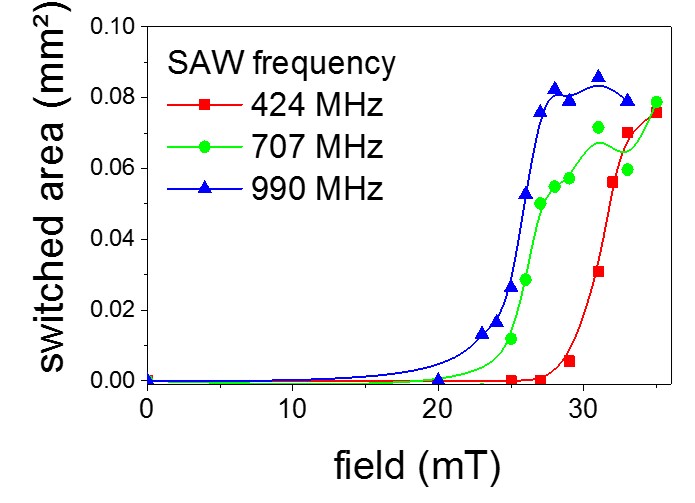}
	\caption{Switched magnetization area in front of IDT \textbf{1} after   a 200ns  SAW burst ($T$=20K,  $\vec{k}_{saw}$//[110],  averaging area is the dotted rectangle in figure \ref{fig:Switch2configs}d). The rf excitation power is adjusted for a constant strain  $ \varepsilon_{xx}$(z=0)$\approx 7 \times 10^{-5}$. Switching at $f_0$=141 MHz is too unefficient at this SAW amplitude to appear on this graph. The field was misaligned by a fraction of a degree, so that at high field, almost 100$\%$ of the area has switched.}
	\label{fig:efficiency}
\end{figure}


 We now show how the efficiency of the switching varies with SAW frequency. This was quantified by averaging the switched area in front of the exciting transducer (dotted rectangle in figure \ref{fig:Switch2configs}d - 20mT). Great care was taken to work at constant SAW amplitude, and constant magnetic field angle.  We show for instance the resulting efficiency plot (figure \ref{fig:efficiency})  for  $\vec{k}_{saw}$//[110], and a constant  surface strain of  $ \varepsilon_{xx}\approx 7 \times 10^{-5}$. Reversal clearly occurs earlier in field with increasing SAW frequency. This is consistent with the increase of the  SAW attenuation with  f$_{saw}$  (shown in the linear regime in figure \ref{fig:SAWFMReffetf}a, and similar behavior at high acoustic powers):  the power is  lost by the SAW to the magnetization, which fuels an efficient reversal. An identical  behavior was found for  $\vec{k}_{saw}$//[1-10].


\subsection{Stationary SAWs: influence of $f_{saw}$ on magnetic patterning period} 



\begin{figure*}
	\centering
	\includegraphics[width=0.99\textwidth]{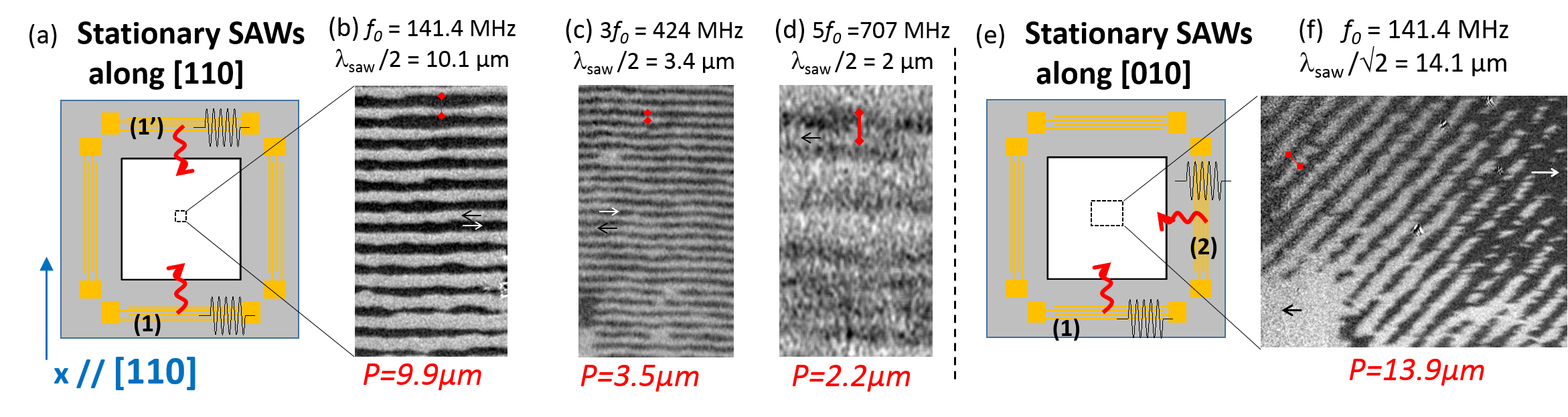}
	\caption{(a) Excitation of stationary SAWs along [110] ($T$=40K).	Close-up of the switched magnetization pattern between IDTs \textbf{1} and\textbf{ 1'} after a 400ns-long SAW burst: (b) $f_0$= 141 MHz ($B$=20mT, $P_{saw}$=13 mW), (c) $3f_0$=424 MHz ($B$=20mT and $P_{saw}$=13 mW), and (d) $5f_0$=707 MHz ($B$=18mT, $P_{saw}$=1.2 mW). The expected $\lambda_{saw}/2$ periodicity (black font) is  calculated using  $\lambda_{saw}=\frac{V_r}{f}$ and $V_r$=2852 m.s$^{-1}$, the experimental one appears in red font and lines. The dashed yellow line is a typical line scan used to make the color maps of figure \ref{fig:statphase}. (e) Excitation of stationary SAWs along [010]. (f) Switched magnetization pattern between IDTs \textbf{1} and \textbf{2} after a 300ns long SAW burst at $f_0$= 141 MHz ($B$=24mT,  image 158$\times$205~$\mu m^2$). For $f_0$ and 3$f_0$ excitation frequencies, the magnification of the microscopy set-up was $\approx$20 (1 pixel=0.295$\mu$m), for 5$f_0$ it was 43 (1 pixel=0.15$\mu$m).  }
	\label{fig:statf}
		\end{figure*}

The data presented up to now has relied solely on \textit{propagating} acoustic waves. Yet the use of \textit{stationary} waves can be relevant for magnetic patterning, as was shown  on FeGa by Li \textit{et al.}\cite{Li2014}  ($f_{saw}$=158 MHz). The mechanism at work was not detailed in this paper, but probably implied  the reduction of domain nucleation/propagation barriers\cite{ThevenardNuc16} rather than precessional switching. Using the different available SAW frequencies of our device, we study the effect of $f_{saw}$ in resonant switching induced by stationary SAWs along [110]. For this,  IDTs \textbf{1} and \textbf{1'} are excited  simultaneously and the magnetization is observed half-way between them (figure \ref{fig:statf}a). The applied field was chosen high enough to have larges patches of homogeneously reversed  magnetization. 

Figures \ref{fig:statf}b-d show a close-up of the center of the sample  after a single 400ns-long burst on each transducer at  $f_0$,   $3f_0$  and $5f_0$. For each of those frequencies, the stationary SAW clearly imprinted a $\lambda_{saw}/2$-periodic pattern to the switched magnetization. This can be explained as follows: for two counter-propagating SAWs of amplitude $A$ and relative phase $\phi$, the stationary wave has an amplitude of the form $A_{110}$=$2Acos(\omega_{saw}t+\phi)cos(\frac{2\pi}{\lambda_{saw}}x+\frac{\phi}{2})$.  Because the SAW burst is much longer than a single period, switching occurs when the SAW amplitude reaches a particular threshold in absolute value\cite{Thevenard2016}, yielding the $\lambda_{saw}/2$ periodicity of  $|A_{110}|$ in the magnetic pattern (figures \ref{fig:statf}b-d).  Note that the micron-resolution of our microscope  prevented us from seeing the stationary pattern at  $7f_0\approx$ 990 MHz  (fringe width $\lambda_{saw}/4$=0.7$\mu m$).

 Likewise, magnetic fringes along [100] can be patterned by exciting simultaneously  two perpendicular SAWs (IDTs \textbf{1} and \textbf{2}, figure \ref{fig:statf}e). This time the amplitude of the  stationary  wave reads $A_{100}=2Acos[\frac{\pi}{\lambda_{saw}}(x+y)-\omega_{saw}t+\phi]cos[\frac{\pi}{\lambda_{saw}}(x-y)+\phi]$. The  $\frac{\lambda_{saw}}{\sqrt{2}}$-periodicity  along [100] expected from $|A_{100}|$ is  evident in  figure \ref{fig:statf}f. A SAW wavefront parallel to the easy axis ($\vec{k}_{saw}$//[110]) is however clearly the more favorable configuration since the resulting magnetic pattern presents a weak magnetostatic cost, the charges being comfortably sent to the extremities of the fringes. This is less true when at least one of the travelling SAW wavefront is perpendicular to the easy axis as in figure \ref{fig:statf}f, where a slight horizontal smudging of the fringes can be seen along [1-10]. 

\begin{figure}
	\centering
	\includegraphics[width=0.49\textwidth]{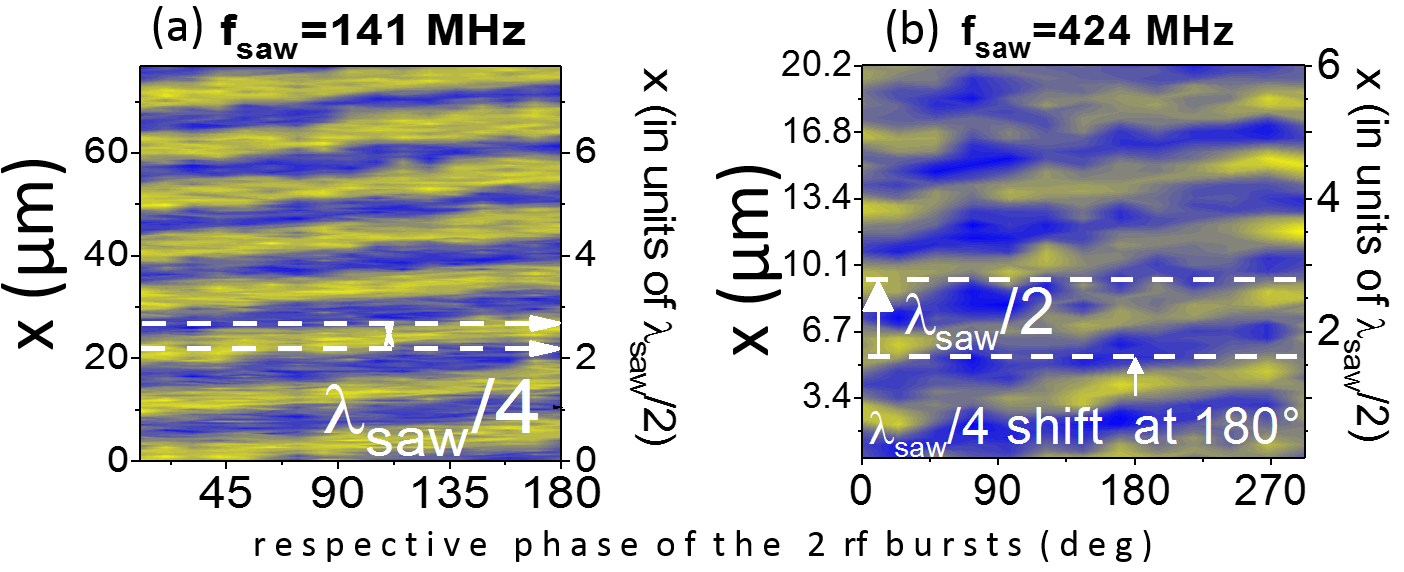}
	\caption{Maps of magnetic contrast along [110] versus relative phase of the 400ns rf bursts on IDTs \textbf{1}/\textbf{1'} (T=40K). The left axis is the x-coordinate along [110] in microns. The right axis is its normalization by $\lambda_{saw}/2$ indicated in figure \ref{fig:statf}. $f_0$= 141 MHz ($B$=20mT, $P_{saw}$=6.3 mW). (b) $3f_0$=424 MHz ($B$=26mT, $P_{saw}$=4mW). }
	\label{fig:statphase}
		\end{figure}

Finally, changing the relative phase of the exciting bursts,  one can also impose the position of the magnetic pattern. This was done for  $f_0$ and $3f_0$ (fringes parallel to [1-10]). The two bursts arriving on IDTs \textbf{1}/\textbf{1'} have been delayed in steps of $\delta \phi$=11.5$^{\circ}$  using a mechanical phase-shifter. Before each of those steps, the sample was reinitialized $+\vec{\textbf{M}}_0$ and a   field  applied along the hard axis. After switching for a given $\phi$ value, the  intensity profile of the fringes along [110] was  plotted, and the low-frequency background removed. A $\delta \phi$-interpolated color map was then created with this data (figure \ref{fig:statphase}). The $\lambda_{saw}/4$ width of a single fringe clearly appears in this graph, as well as the $\lambda_{saw}/4$ shift (half-period of the pattern) for a 180$^{\circ}$ dephasing. An important consequence ensues from this experimental result: while resonant switching using stationary SAWs is probably not the most adapted to imprint very small magnetic patterns, the wave-like nature of the switching implies that a very precise, nanometric positionning of these patterns could easily be implemented, an appealing proprety for reconfigurable magnetic memories.

\section{Compatibility of multi-domain formation with precessionnal switching mechanism} 


Now that a clear experimental demonstration of SAW-driven switching of in-plane magnetization has been presented, we focus on the final multi-domain configuration that was obtained (figure \ref{fig:Switch2configs}), and whether it is compatible or not with a resonant switching mechanism. 

In the simplified  macrospin theory of  precessional switching,  when  starting from a given initial magnetization $\pm\vec{\textbf{M}}_0$, switching to $\mp\vec{\textbf{M}}_0$ should only be possible if the duration of the effective pulse field is an odd-multiple of half the precession frequency under the applied bias field\cite{Thevenard2013}. This feature of precessional switching was  beautifully demonstrated  experimentally   on small metallic devices (typically   100nm to 10$\mu m$-wide ellipses) for very short excitation pulses (electric\cite{Shiota2011} or magnetic fields\cite{Papusoi2009}, or current\cite{Vaysset2011}). Beyond a few nanoseconds duration and for larger samples however, coherence was destroyed. Spatial and temporal dephasing of the precession was due to  thermal effects (via the stochastic nature of the begining of the precession),  or domain wall creation\cite{Hiebert2002,Devolder2008,Vaysset2011,Balestriere2011a} (for bigger structures), rather than chaotic magnetization trajectories\cite{Devolder2005}. This lead  to the switched/unswitched probability rapidly converging towards 50$\%$ with excitation pulse duration. We expect  similar phenomena to occur when driving precessional switching acoustically, with the dephasing caused both by the spread in anisotropy constants in the magnetic layer and the relatively long duration of the excitation pulses.

 \begin{figure*}
	\centering
	\includegraphics[width=0.98\textwidth]{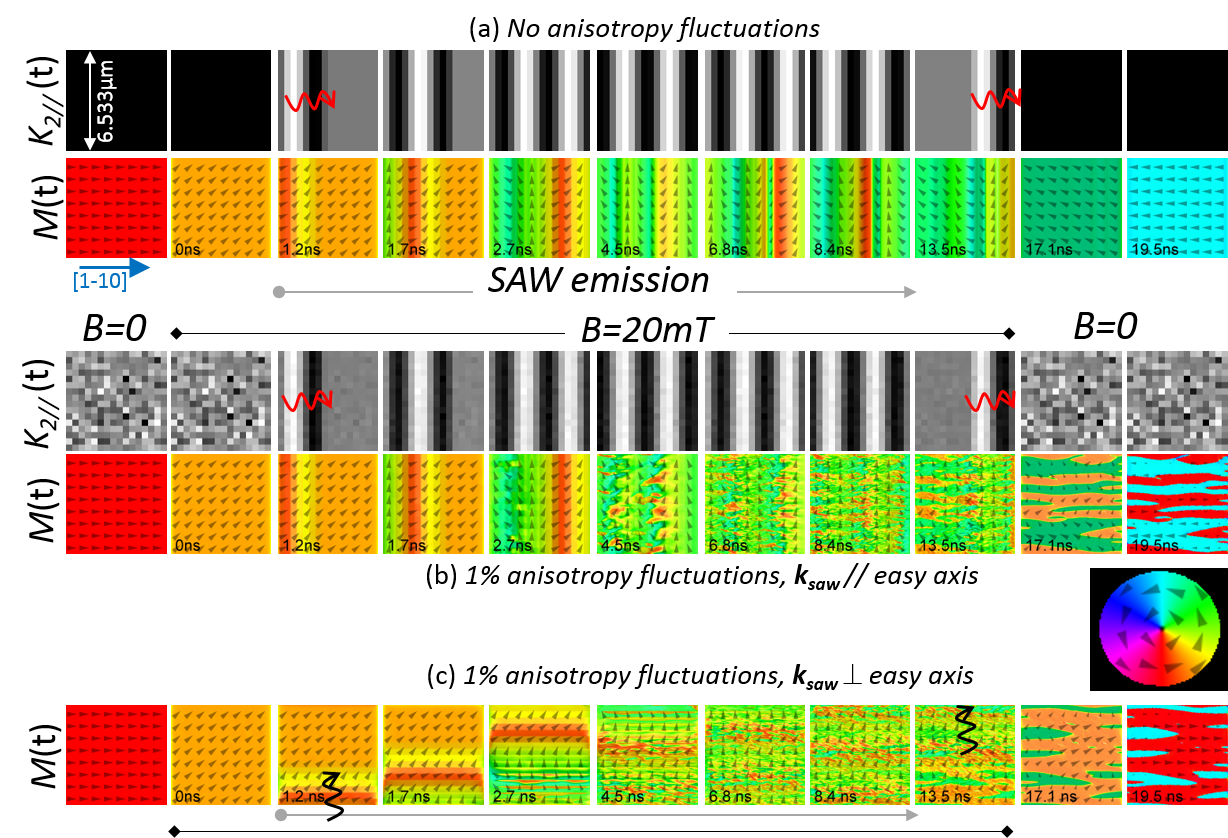}
	\caption{Micromagnetic simulations of SAW-driven resonant  magnetization reversal ($B$=20 mT,  other parameters indicated in Appendix B, magnetization color wheel in the bottom right-hand side). The simulation window is 2.3$\lambda_{saw}$ wide. The SAW is emitted from the left between $t$=0.1 and 12.12ns (12T$_{saw}$). The field is turned on between $t$=0 and 17ns (see full movies in the supplementary information). (a) No anisotropy dispersion for the static $K_{2||},K_{c}$, and  $\vec{k}_{saw}||$[1-10]. (b) 1$\%$ anisotropy fluctuations for the static $K_{2||},K_{c}$, and $\vec{k}_{saw}||$[1-10]. The top panel is a normalized map of the time- and space- dependent uniaxial anisotropy modulated by the SAW. The bottom panel is the state of the magnetization at the corresponding time.  (c) Same as (b), but with $\vec{k}_{saw}||$[110]. }
	\label{fig:mumaxtime}
		\end{figure*}

\begin{figure}
	\centering
	\includegraphics[width=0.48\textwidth]{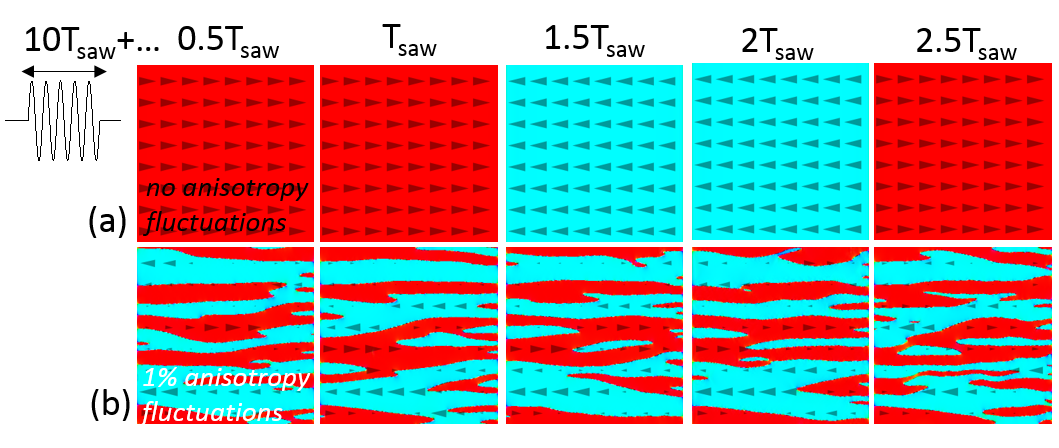}
	\caption{Effect of the SAW pulse duration $\tau_{saw}$=10-12.5$T_{saw}$  on the final remanent state, starting from $+\vec{\textbf{M}}_0$  ($\vec{k}_{saw}||$[1-10], $B$=20mT, other parameters indicated in Appendix B). Red (resp. blue) areas have not switched (resp. switched) to  $-\vec{\textbf{M}}_0$: (a) No anisotropy dispersion for $K_{2||},K_{c}$:  the switching is 2$T_{saw}$-periodic as expected from the macrospin approach. (b) 1$\%$ anisotropy fluctuations for $K_{2||},K_{c}$.  A minor dispersion in the anisotropy creates a multi-domain structure that weakly depends on the SAW pulse duration.}
	\label{fig:mumax}
		\end{figure}

Elegant  analytical solutions  have been derived for field and STT-driven precessional reversal of spin valves or magnetic tunnel junctions\cite{Mayergoyz2004,Devolder2004,Devolder2004a,DAquino2016}. In order to keep calculations tractable, they are however often limited to idealized ellipsoid geometries.  Being moreover macrospin calculations, they do not take into account exchange interaction and dynamic demagnetizing effects, which are expected to have a central role in the final switched domain configuration of our system.  We therefore implement instead micromagnetic  MuMax$^3$ simulations \cite{Vansteenkiste2014}. We first assumed no dispersion of the native  uniaxial anisotropy constant $K_{2||}$.  The effect of the propagating SAW on the system is modelled as: $K_{2||}(x,t)$=$K_{2||}+\Delta K_{2||}cos(k_{saw}x-\omega t)$  (see Appendix B for technical details on the simulations, and figure \ref{fig:mumaxtime}a-top for snapshots of the anisotropy). $\Delta K_{2||}$ is proportional to the SAW amplitude and to the magneto-elastic coefficient $A_{2xy}$.   Starting from a uniform $+\vec{\textbf{M}}_0$ configuration, we apply a 20 mT magnetic field along the hard axis. Figure \ref{fig:mumaxtime}a-bottom shows selected snapshots of the time  evolution of the magnetization (see the full "M(t) no dispersion.avi" movie in the supplementary information). As expected from macrospin calculations\cite{Thevenard2013}, the magnetization follows a wide angle precession around the vertical (hard axis) direction. Once the SAW emission is stopped ($t$=12.12ns), the acoustic pulse finishes propagating across the simulation window, and the magnetization returns to a small damped precession around its final equilibrium position. For  $\tau_{saw}$=12$T_{saw}$, the final state after removal of the field is a fully switched \textit{uniform} magnetization $-\vec{\textbf{M}}_0$, even though exchange and demagnetization energy costs have been taken   into account. Moreover,  for this magnetic field/SAW amplitude combination guarantying large amplitude oscillations of  $\vec{\textbf{M}}(\vec{r},t)$,  the final state can be chosen   by  how many multiples of $T_{saw}$ have elapsed during $\tau_{saw}$ (figure  \ref{fig:mumax}a).  The remanent state  depends quite subtly on the detuning between the SAW frequency and the  precession frequency \cite{Thevenard2016}.

When 1$\%$ fluctuations are included  in  the magneto-elastic constants, a very different behavior is observed (figures \ref{fig:mumaxtime}b,c and  full "\textit{M(t) dispersion-SAW along easy/hard axis.avi}" movies in the supplementary information). After a few SAW periods, the precession amplitude starts to vary across the SAW wavefront, introducing a progressive dephasing that is amplified as the SAW  progresses. 
The spatial variations of the magneto-elastic constants reflect on the local efficiency of the SAW rf field. This triggers the nucleation of pockets of anomalies, which  the SAW propagation then streches out into  characteristic filaments. When the field is switched off ($t$=17 ns), exchange and magnetostatic energies govern the formation of  clear micron-sized fully switched (blue) or unswitched (red) domains. The system has limited the formation of charged transverse domain walls along  [110], preferring\cite{Thevenard2017} the less costly N\'eel walls along [1-10]. This results in filamentation along the the easy axis. Varying the SAW pulse duration changes only weakly the final state (figure \ref{fig:mumax}b), contrary to the case when no anisotropy fluctuations are included. The exact same simulation can be run  with  $\vec{k}_{saw}||$[110] (figure \ref{fig:mumaxtime}c). In that case, the final domains are larger and less filamented than for the $\vec{k}_{saw}||$[1-10]  case, since the SAW wave-front evolves parallel to the magnetic easy axis, a much more confortable situation from the point of view of magneto-statics.

  We need to underline that the code implemented for the MuMax simulations imposes an identical spatial grid (here with a super-cell of 408nm) to describe the SAW propagation and the anisotropy fluctuations, so that large scale anisotropy fluctuations (e.g region size of a few microns) cannot be combined with a fine description of the SAW (e.g region size below $\lambda_{saw}/20$).  This is possibly why the   characteristic micron-width of the filaments found in the simulation is smaller than the one observed experimentally (around 10-20$\mu$m). A full study of the critical dispersion scale to SAW wavelength ratio leading to filamentation is therefore out of the scope of the present paper.  Yet, the simulations reproduce qualitatively very well  three  experimental features  (figure \ref{fig:Switch2configs}): (i) the final state is not uniformly magnetized, but a collection of  $+\vec{\textbf{M}}_0$ and  $-\vec{\textbf{M}}_0$ domains. For  $\vec{k}_{saw}||$[1-10], these take the form of filaments running roughly parallel to $\vec{k}_{saw}$, (ii) the final state is very weakly dependent on the SAW pulse duration, (iii) the switched domains are broader and less filamented when the SAW travels along the hard axis rather than along the easy axis.



\section{Conclusion}

We have presented a comprehensive investigation of resonant magnetic switching by a surface acoustic wave with a special emphasis on the role of the SAW frequency and the interplay between the SAW wavevector direction and the easy magnetic axis. The main results are the demonstration that : (i) Resonant reversal of in-plane magnetization is possible   using large amplitude Rayleigh waves travelling along the easy or hard magnetic axis. The main difference between the two geometries lies in the aspect ratio  of the reversed domains, with the magnetization clearly trying to reduce its magnetostatic energy by forming jagged or pointy domains along the easy  axis.  (ii) The reversed domains are  several hundreds of $\mu m^{2}$, 
 much larger than previous demonstrations of STT- or field-driven precessional switching. This is also  in stark contrast with SAW-induced switching in  out-of-plane magnetized (Ga,Mn)(As,P) on which both the large spread of magnetic anisotropy and the natural tendency  to self-organization into up/down domains meant that the switching occurred over small, sub-micron areas\cite{Thevenard2016}. (iii) When the SAW wavefront lies parallel to the easy axis, the switched domains are large enough for  periodic magnetic patterns to be carved out in stationary geometry,  with a sub-micron positionning precision. There is  no  restriction to the minimum magnetic  period that can be reached, save that of being able to excite high frequency SAWs (and to be able to observe sub-micron features). (iv) Micromagnetic simulations clearly confirm that  the formation of multi-domains and the lack of pulse-length dependence of the final switched state are both compatible with a resonant precessional mechanism.

   The limitations of this switching scheme have also been evidenced. The final pattern appeared very sensitive to field alignment and spatial variations of the anisotropy. This has also been one of the bottlenecks of precessional switching of MTJ or spin-valves, for which the asymmetry in the switching probabilites between parallel and anti-parallel configurations was due to the current-generated Oersted field or the stray field from the polarizing/analyzing layers\cite{Lee2011a,Liu2012a,Devolder2008,Park2013}.   Several strategies were implemented in these systems to bypass this issue, such as the use of a synthetic antiferromagnet perpendicular polarizer to limit spurious stray fields. In the SAW switching scheme, a possible route would be to engineer the magnetic anisotropy to have the zero-field precession frequency close to f$_{saw}$.
   
   The long rise time of the SAW pulse - imposed by the large number of teeth required to excite efficiently the chosen frequency - is also likely detrimental to the determinism of the switching. The optimal design for SAW switching would therefore include a single digit-pair transducer (very short transient regime), with a SAW emission centered around a high frequency (high efficiency), and a uniaxial layer  structured into sub-micron devices. This should allow  to chose  the final switched state whilst maintaining the advantages of  propagating/stationary wave to adress selectivally a structure spatially.       
    


\section{acknowledgements}

 This work has been supported by the French Agence Nationale de la Recherche (ANR13-JS04-0001-01). The authors also acknowledge David Hrabovsky  of the MPBT (physical properties - low temperature) facility of Sorbonne Universit\'e for  technical support.

\section*{APPENDIX}

\subsection{A. Working at constant SAW amplitude}

 We have recently developed two precise techniques to  estimate the SAW amplitude, using X-ray diffraction\cite{Largeau2016} or a vector network analyser\cite{Camara2016}. The former is difficult to implement at cryogenic temperatures. The latter relies on  measurements of the $S$ parameters  and modelling using the Coupling-of-Modes theory. It could however not be implemented here:  (i) no "dip" was observed in the $|S_{ii}|^2$ reflection coefficients, and (ii) working at cryogenic temperatures made it quite complicated to evaluate precisely the contribution of the fixtures, an indispensable step in this approach. For this reason, we proceeded to a gross estimate of the acoustic power as follows.  Exciting IDT \textbf{i} with an rf voltage of amplitude $U_{exc}$, we detect an echo of amplitude $U_{det}$ on  the opposite transducer \textbf{i'}. The incident electrical power $P_{exc}$ is first measured on a matched load (50 $\Omega$). The echo amplitude is then measured
on a 50 $\Omega$ load and converted to an electrical power $P_{det}$. Assuming an equal transduction coefficient $\beta$ for
both IDTs, we get $P_{det}$ = $\beta^2 P_{exc}$, from which $\beta$ is computed. Finally, the acoustic power is estimated
as  $P_{saw}$=$\beta P_{exc}$.  The expression of the Poynting vector then yields the numerical value of the strain\cite{royer2000elastic}. Note that this was performed at low temperature and in zero field (i.e far from any magnetic resonance capable of damping the SAW amplitude, see figure \ref{fig:SAWFMReffetf}a). For the device presented here, the transduction insertion loss amounted to   -25dB to -30dB, depending on the frequency, the IDT and the temperature.

\subsection{B. Micromagnetic simulations parameters}

 The size of the simulation is 6533$\times$6533$\times$50nm ($L_x\times L_y \times L_z$) with 512$\times$512$\times$4 cells. Periodic boundary conditions are applied in the direction perpendicular to $\vec{k}_{saw}$.   The following magnetic parameters were used: $M_s$ = 50 kA.m$^{-1}$, Gilbert damping $\alpha$=0.1, $K_{2\bot}$= -8133 J.m$^{-3}$,$K_{2||}$=800 J.m$^{-3}$,  exchange constant\cite{Shihab2015} $A_{ex}$ = $10^{-13}$ J.m$^{-1}$, cubic anisotropy $K_c$=300 J.m$^{-3}$. MuMax on our OS does not accomodate two uniaxial anisotropies, so  only  the in-plane uniaxial anisotropy was considered. Discarding   the out-of-plane anisotropy $K_{2\perp}$ effectively lowers the precession frequency and affects the in-plane and out-of-plane torques of the SAW on the magnetization. In this respect,  quantitative comparisons  between these simulations and the experiments should be made with great care. Acoustic parameters were taken as: $f_{saw}$=990 MHz and $V_r$=2852 m.s$^{-1}$, which yields $\lambda_{saw}$=2.8 $\mu$m. Anisotropy fluctuations were implemented by dividing the simulation window into 16$\times$16 regions of coordinates $\vec{r}_i=(x_i,y_i)$, in which  small random  anisotropy variations proportionnal to $R$ could be added to the homogenous static  anisotropies $K_{2||}$ and  $K_c$. Using those same 16$\times$16 regions, the effect of the propagating SAW was modelled by adding  a time- and space varying component $\Delta K_{2||}(\vec{r}_i,t)$ (see "Ku-dispersion.avi"  movie in the supplementary material). Each region is 408nm$\approx \frac{\lambda_{saw}}{7}$ wide, a satisfyingly small fraction of the SAW wavelength. For $\vec{k}_{saw}$ along the easy axis, the final  uniaxial anisotropy reads: $K_{2||}(\vec{r}_i,t)$=$K_{2||}$(1+randNorm($\vec{r}_i$)R)+  $\Delta K_{2||}$(1+randNorm($\vec{r}_i$)R)cos($k_{saw}x_i-\omega t)H(t-t_0+\frac{x_i}{V_r})H(-(t-t_0-\frac{x_i}{V_r}-\tau_{saw})$, with $ \Delta K_{2||}$=200 J.m$^ {-3}$,  $\tau_{saw}$ = 10-12.5ns$\approx$10-12.5$T_{saw}$ the SAW pulse duration,  $t_0$=100~ps, and $R$=0 or 0.01 depending on whether anisotropy fluctuations are taken into account or not.  The field is turned off at $t_{f}$=17ns, a few ns after the SAW has completely left the simulation window. The simulation is then run for an extra 4ns, i.e  over twice typical damping times in this material\cite{Shihab2015}.

\bibliographystyle{apsrev4-1}

%

\end{document}